\documentstyle[twoside,fleqn,espcrc2,epsf]{article}

\newcommand{\AmS}{{\protect\the\textfont2
  A\kern-.1667em\lower.5ex\hbox{M}\kern-.125emS}}

\newcommand{\be}{\begin{equation}}
\newcommand{\ee}{\end{equation}}
\newcommand{\bea}{\begin{eqnarray}}
\newcommand{\eea}{\end{eqnarray}}

\normalsize
\title{
Light hadron spectroscopy with two flavors 
of $O(a)$ improved dynamical quarks
\thanks{Talk presented by T.~Kaneko}}

\author{
JLQCD Collaboration:
S.~Aoki\address{Institute of Physics, University of Tsukuba, 
                Tsukuba, Ibaraki 305-8571, Japan},
R.~Burkhalter$^{\rm a,}$\hspace{-1mm}
             \address{Center for Computational Physics, 
             University of Tsukuba, Tsukuba, 
             Ibaraki 305-8577, Japan}, 
M.~Fukugita\address{Institute for Cosmic Ray Research,
            University of Tokyo, Kashiwa 277-8582, Japan}, 
S.~Hashimoto\address{High Energy Accelerator Research Organization
                     (KEK), Tsukuba, Ibaraki 305-0801, Japan},
K-I.~Ishikawa$^{\rm d}$,
N.~Ishizuka$^{\rm a,b}$, 
Y.~Iwasaki$^{\rm a,b}$,   K.~Kanaya$^{\rm a,b}$, 
T.~Kaneko$^{\rm d}$,      Y.~Kuramashi$^{\rm d}$, 
M.~Okawa$^{\rm d}$,       
T.~Onogi\address{Department of Physics, Hiroshima University
                 Higashi-Hiroshima, Hiroshima 739-8526, Japan}, 
S.~Tominaga$^{\rm b}$,    N.~Tsutsui$^{\rm d}$,
A.~Ukawa$^{\rm a,b}$,     N.~Yamada$^{\rm d}$,  and 
T.~Yoshi\'e$^{\rm a,b}$.
} 

\begin{document}

\begin{abstract}
  We report on our study of light hadron spectrum and quark
  masses in QCD with two flavors of dynamical quarks. 
  Simulations are made with the plaquette gauge action and 
  the non-perturbatively $O(a)$ improved Wilson quark action.
  We simulate 5 sea qaurk masses corresponding to 
  $m_{\rm PS}/m_{\rm V} \!  \simeq \! 0.8$--0.6 at $\beta=5.2$
  on $12^3 \times 48$, $16^3 \times 48$ and 
  $20^3 \times 48$ lattices.
  A comparison with previous calculations in quenched QCD
  indicates sea quark effects in meson and quark masses.  
\end{abstract}

\maketitle
\setcounter{footnote}{0}

\section{Introduction}
One of the major goals of lattice QCD simulation is to
confirm the validity of QCD as the theory of strong
interaction in the low energy region by comparing its
prediction for the hadron spectrum to experiment. 
Because of huge computational demand to perform simulations
with dynamical quarks, however, many works have been forced
to neglect effects of dynamical sea quarks.
In this quenched approximation, the CP-PACS collaboration
found some deviation in the light hadron spectrum from
experiment\cite{CP-PACS.Quenched}, which motivates us to
investigate sea quark effects by performing simulations
of QCD with the realistic number of dynamical quark flavors.

The JLQCD collaboration started numerical simulations of
dynamical QCD on a supercomputer Hitachi SR8000 model F1, 
which is newly installed at KEK on March 2000.
It has a 1.2 TFlops of peak performance and 
provides about 400 sustained GFlops for our simulation code.
We carry out simulations of QCD with two flavors 
of dynamical quarks and investigate 
the chiral extrapolation and finite size effects 
in the light hadron spectrum,
as a step toward studies of QCD with 2+1 flavors
(i.e. $u$, $d$ plus $s$ quarks).

\section{Simulations}
We study QCD with two degenerate flavors of dynamical quarks, 
which are identified with $u$ and $d$ quarks;
the strange quark is treated in the quenched approximation.
We employ the $O(a)$ improved quark action\cite{clover-action} 
with the clover coefficient $c_{\mathrm{SW}}$ determined non-perturbatively
by the ALPHA collaboration\cite{NP-csw}.
Simulations are performed at $\beta\!=\!5.2$ and 
$c_{\mathrm{SW}}\!=\!2.02$, for which $a^{-1}$ evaluated at
the physical mass of dynamical $ud$ quarks 
is approximately 2~GeV and
hence scaling violation is not expected to be too large.
We simulate 5 sea quark masses in the range
$m_{\mathrm{PS}}/m_{\mathrm{V}}\! \simeq \! 0.8$--0.6.
Since finite size effects could be more important with
dynamical quarks, we perform simulations on three lattices
with different spatial sizes, 
$12^3 \times 48$, $16^3 \times 48$ and $20^3 \times 48$.
We have accumulated 3000 thermalized HMC trajectories on
the $16^3 \times 48$ lattice, while simulations on other lattices
are in progress.
Other details of simulation parameters are summarized 
in Table~\ref{tab:param}. 

\begin{table}[t]
\vspace*{-5mm}
\setlength{\tabcolsep}{0.2pc}
\caption{
  Simulation parameters.
  We also list the lattice spacing $a_{r_0}$ 
  fixed by $r_0\!=\!0.49$~fm for $16^3 \times 48$ lattices.
}
\begin{tabular}{lllll}
\hline
 lattice   & $K_{\mathrm{sea}}$ & \#traj. & $m_{\mathrm{PS}}/m_{\mathrm{V}}$ 
& $a_{r_0}$~[fm]\\
\hline
$16^3{\times}48$     &  0.1340   & 3000 &   0.802(4)  & 0.1288(10) \\
                     &  0.1343   & 3000 &   0.781(6)  & 0.1201(10) \\
                     &  0.1346   & 3000 &   0.743(6)  & 0.1127(7) \\
                     &  0.1350   & 3000 &   0.714(9)  & 0.1084(7) \\
                     &  0.1355   & 3000 &   0.596(18) & 0.1006(5) \\
                     \cline{2-4}
\hline
$12^3{\times}48$     &  0.1346   & 3000 &   0.735(9)  & -- \\  
                     &  0.1350   & 3000 &   0.695(15) & -- \\
\hline
$20^3{\times}48$     &  0.1346   & 2000 &   0.756(8)  & -- \\  
                     &  0.1350   & 2800 &   0.706(9)  & -- \\
\hline
\end{tabular}
\label{tab:param}
\vspace{-4mm}
\end{table}

\section{Static quark potential}
\begin{figure}[t]
\vspace{-12mm}
\begin{center}
\leavevmode
\epsfxsize=7.3cm
\epsfbox{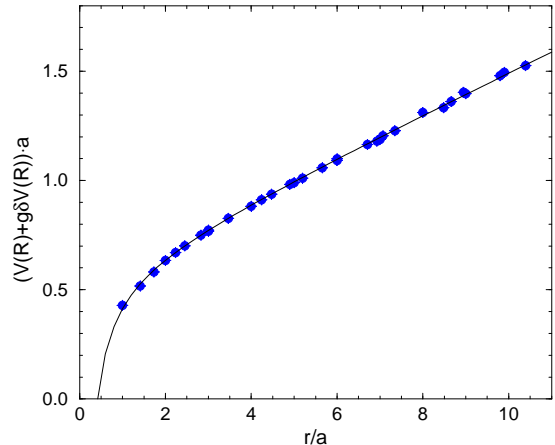}
\end{center}
\vspace{-16mm}
\caption{
  Corrected static quark potential 
  $V(r)\!+\!g \cdot \delta V(r)$ at $K_{\mathrm{sea}}\!=\!
  0.1340$ on $16^3 \times 48$ lattice. 
  The solid line represents the fitting curve (\ref{eq:pot}).
}
\label{fig:potential}
\vspace{-7mm}
\end{figure}

We calculate the static quark potential on the
$16^3 \times 48$ lattice using the smeared Wilson loops.
Since we do not observe any signal of flattening
of the potential due to string breaking,
potential data are parameterized with the form
\begin{eqnarray}
  V({\bf r}) = V_0 - \alpha/|{\bf r}| 
             - g \cdot \delta V({\bf r})
             + \sigma \cdot |{\bf r}|, 
  \label{eq:pot}
\end{eqnarray}
where $\delta V({\bf r})\!=\!(G({\bf r})-1/|{\bf r}|)$
represents a correction to the short-distance Coulomb term 
calculated 
with the lattice gluon propagator\cite{potential}.
Figure~\ref{fig:potential} shows a plot obtained for the
heaviest sea quark mass, where we find
that deviation of the potential data from the fit curve is
at most 1\% in the fitting range of ${\bf r}$.
We determine the Sommer scale $r_0$ \cite{r0} from the
parameterization (\ref{eq:pot}).
The lattice spacing determined with the condition 
$r_0\!=\!0.49$~fm is listed in Table~\ref{tab:param}.

\section{Light meson mass measurement}

We calculate light hadron correlators for 6 values of the
valence quark masses in the range
$m_{\rm PS}/m_{\rm V} \! \simeq \! $ 0.80--0.50 for each sea quark
mass. 
Measurements have been completed for the $16^3 \times 48$
lattice, while only the degenerate case 
($K_{\mathrm{val}}\!=\!K_{\mathrm{sea}}$) is measured on 
$12^3 \times 48$ and $20^3 \times 48$.
Most of the analysis results discussed in this talk are based on the 
$16^3 \times 48$ lattice data except for the finite size
effect in the following.

\begin{figure}[t]
  \vspace{-10mm}
  \begin{center}
    \leavevmode
    \epsfxsize=7.3cm
    \epsfbox{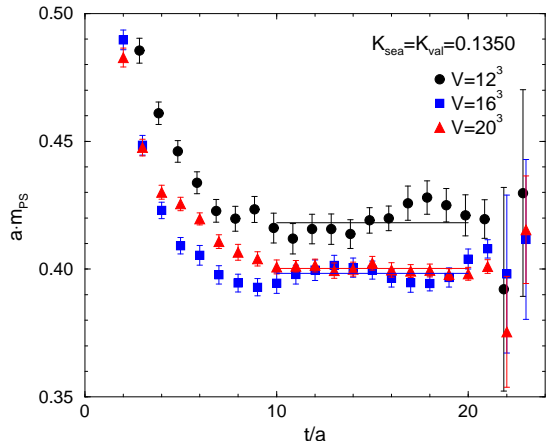}
  \end{center}
  \vspace{-16mm}
  \caption{
    Effective masses for the pseudo-scalar meson at 
    $K_{\rm sea}\!=\!0.1350$ on three different volumes.
    Solid lines represent the central value of the fit
    result.
    } 
  \label{fig:FSE-em}
  \vspace{-4mm}
\end{figure}

\begin{figure}[t]
  \vspace{-13mm}
  \begin{center}
    \leavevmode
    \epsfxsize=7.3cm
    \epsfbox{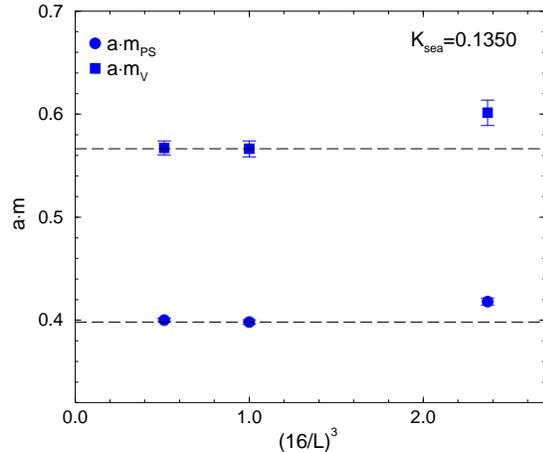}
  \end{center}
  \vspace{-16mm}
  \caption{
    Fit results of pseudo-scalar (circles) and vector
    (squares) meson masses as a function of 
    $(16/L)^3$.
    Results on $V\!=\!16^3$ are shown also with
    dashed lines for a guide of eye.
    }
  \label{fig:FSE-fitted}
  \vspace{-6mm}
\end{figure}

Figure~\ref{fig:FSE-em} shows the effective mass of
degenerate ($K_{\mathrm{val}}=K_{\mathrm{sea}}$) 
pseudo-scalar meson at the second lightest sea quark mass.
Data are shown for three different spatial volumes 
$12^3$, $16^3$ and $20^3$.
We find that the fitted masses 
on the two larger lattices are in good agreement with each other
both for the pseudo-scalar and vector mesons, 
as shown in  Figure~\ref{fig:FSE-fitted}.
This is also observed for other lattices with heavier sea quarks.
This suggests that finite size effects are already small for
the 16$^3$ lattice.
We are currently extending similar measurements to the
lightest sea quark $K_{\mathrm{sea}}\!=\!0.1355$, for which
finite size effects are expected to be more important.

\section{Sea quark effect in meson masses}

\begin{figure}[tb]
  \vspace{-12mm}
  \begin{center}
    \leavevmode
    \epsfxsize=7.3cm
    \epsfbox{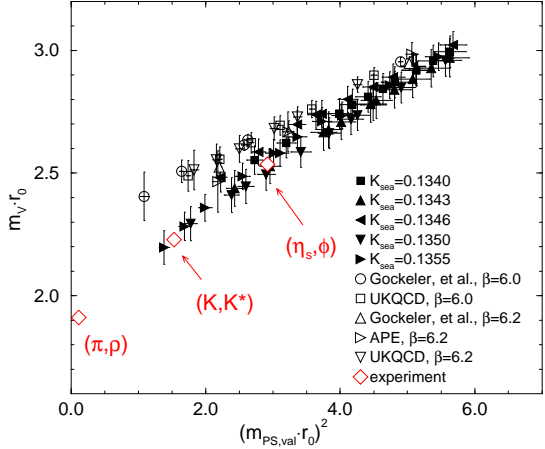}
  \end{center}
  \vspace{-15mm}
\caption{
  $(m_{\mathrm{V}} r_0)$ vs $(m_{\mathrm{PS}} r_0)^2$
  in two flavor QCD (filled symbols).
  Open symbols represent results in quenched QCD 
  in Refs.~\cite{qNP-results}.
  Experimental values are plotted with open diamonds
  using $r_0\!=\!0.49$~fm.
  }
\label{fig:meson-finite_msea}
\vspace{-5mm}
\end{figure}

Figure \ref{fig:meson-finite_msea} shows the vector meson
mass as a function of the pseudo-scalar meson mass
squared. 
Since the effective lattice spacing decreases as the sea
quark mass becomes smaller, which complicates the chiral
extrapolation if made using the lattice unit, we plot the meson
masses normalized by a physical quantity $r_0$ measured for
each sea quark mass.
We also plot the quenched data \cite{qNP-results} in
Figure~\ref{fig:meson-finite_msea}, where we clearly see an
indication of sea quark effect,
i.e. slope in two-flavor QCD is significantly larger than
that in quenched QCD.
This leads to a larger hyperfine splitting of the strange
meson as we shall see later. 
We also find that our results at different sea quark masses
lie almost on one curve.
This indicates that the sea quark mass dependence 
in this quantity is not large
in our simulated region of $K_{\rm sea}$ and 
more precise calculation is needed, particularly for 
the vector mesons, to see it clearly.

\section{Chiral extrapolation}
We employ two different strategies
for the chiral extrapolation.
In our main strategy, which we call method (A), 
we use meson and quark masses normalized by $r_0$
throughout the analysis. 
For instance, the pseudo-scalar meson mass squared is fit to
the form 
\begin{equation}
  (m_{\mathrm{PS}} r_0)^2 
  = A_s (m_{\mathrm{sea}} r_0)
  + A_v (m_{\mathrm{val}} r_0),
\end{equation}
where $m_{\mathrm{sea}}$ and $m_{\rm val}$ are respectively
sea and average valence quark masses defined through the 
vector Ward identity (VWI) relation
$m_{q,\mathrm{VWI}}=(1/K-1/K_c)/2$.
With this method, data is described very well by a
simple linear fit.
In the other strategy, method (B), we make the chiral
extrapolation using meson and quark masses in the lattice unit. 
In this case, significant curvature is found in fits 
for both pseudo-scalar and vector meson masses, and therefore
we have to introduce quadratic terms in the fitting function.
We find that both fits give consistent results, although the
fit error is generally larger for the method (B).

In each method, the physical quark masses $m_{ud}$ and $m_{s}$
are determined by tuning meson mass ratios to their
experimental values.
We use $m_{\pi}$ and $m_{\rho}$ to fix $m_{ud}$ and the
lattice spacing. The strange quark mass $m_{s}$ is
determined from either $m_{K}/m_{\rho}$ or
$m_{\phi}/m_{\rho}$.

\begin{figure}[t]
  \vspace{-12mm}
  \begin{center}
    \leavevmode
    \epsfxsize=7.3cm
    \epsfbox{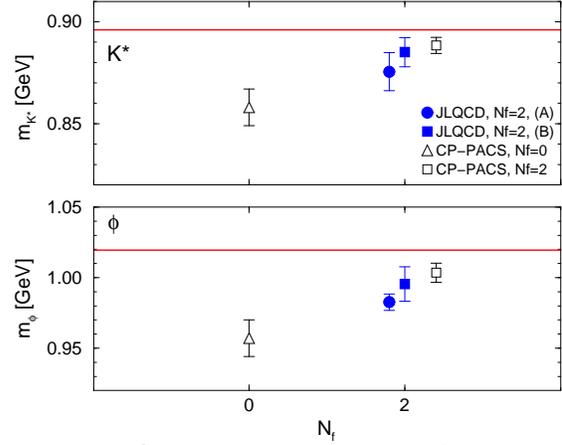}
  \end{center}
  \vspace{-16mm}
  \caption{
    Strange meson masses with $K$-input.
    Filled circles and squares are our full QCD results 
    using method (A) and (B), respectively.
    Open triangles and squares are CP-PACS's result 
    in quenched\cite{CP-PACS.Quenched} 
    and full QCD\cite{CP-PACS.full}. 
    Horizontal lines show experimental values.
    }
  \label{fig:meson}
  \vspace{-6mm}
\end{figure}

Our results for the strange vector meson masses ($K^*$ and
$\phi$), with pseudo-scalar ($K$) used to fix the strange
quark mass, are shown in Figure~\ref{fig:meson} by filled
circles (method (A)) or by filled squares (method (B)).
For comparison, we also plot the CP-PACS data in
the quenched approximation \cite{CP-PACS.Quenched} and
the recent two-flavor result \cite{CP-PACS.full}.
A trend that the sea quark effect pushes up the vector meson
masses toward the experimental values
is seen in this plot.

\section{Strange quark mass}
The bare strange quark mass is calculated using either of
two definitions, i.e. one from VWI and the other from the 
axial vector Ward identity (AWI), 
$2m_{q,\rm AWI} \!= \!
   \langle \partial_4 A_{4}(t) P(0)^{\dag} \rangle
  /\langle P(t) P(0)^{\dag} \rangle$, 
where 
$A_{\mu}(t)$ and $P(t)$ are the axial vector current
and pseudo-scalar density.
We use the improved axial vector current 
$A_{\mu}^{(imp)}(t) \! = \! A_{\mu}(t) 
                         +c_A \partial_{\mu} P(t)$,
with $c_A$ calculated at one-loop \cite{Improved-A}.
The continuum $\overline{\rm MS}$ quark mass is obtained
using one-loop matching \cite{Improved-A,Zfactor} at scale
$\mu\!=\!1/a$ and evolved to $\mu\! =\! 2$~GeV with 3-loop $\beta$
function\cite{3loop-running}.

Our results for the strange quark mass using the method (A)
are 
\begin{equation}
  m_{s}(2 \mathrm{GeV}) = 
  \left\{
    \begin{array}{ll}
      \mbox{94(2)~MeV\ } & \mbox{(VWI)},\\
      \mbox{88(3)~MeV\ } & \mbox{(AWI)},  
    \end{array}
  \right.
  \label{eq:m_s_K-input}
\end{equation}
with $K$ used as input, or
\begin{equation}
  m_{s}(2 \mathrm{GeV}) =  
  \left\{
    \begin{array}{ll}
      \mbox{109(4)~MeV\ } & \mbox{(VWI)}, \\
      \mbox{102(6)~MeV\ } & \mbox{(AWI)}, 
    \end{array}
  \right.
  \label{eq:m_s_phi-input}
\end{equation}
with $\phi$ used as input.
The method (B) gives consistent results.

Only the statistical errors are shown in
(\ref{eq:m_s_K-input}) and (\ref{eq:m_s_phi-input}).
The systematic error is more significant, as indicated by the
disagreement between VWI and AWI, and the difference
between (\ref{eq:m_s_K-input}) and
(\ref{eq:m_s_phi-input}), although the latter is smaller
than in the quenched results\cite{CP-PACS.Quenched}. 
One of the most important sources of the systematic error is 
the use of the one-loop perturbative $Z$ factor, for which
a naive order counting gives $O(5\%)$.
In addition, the discretization error of $O(a^2)$ or the
quenching effect of the strange sea quark could also be
important. 
A recent two-flavor simulation of CP-PACS predicted 
$m_s(2 {\rm GeV})=88^{+4}_{-6}$ MeV or $90^{+5}_{-11}$ MeV
in the continuum limit with the $K$ or $\phi$ used as
input \cite{CP-PACS.full}.

\vspace{3mm}
This work is supported by the Supercomputer project No.54
(FY2000) of High Energy Accelerator Research Organization
(KEK), and also in part by the Grant-in-Aid of the Ministry
of Education 
(Nos.~10640246, 10640248, 11640250, 11640294, 11740162,
12014202, 12640253, 12640279 and 12740133).
K-I.I, T.K and N.Y are supported by the JSPS Research Fellowship.

\end{document}